\documentclass[twocolumn]{aastex61}

\usepackage{url}
\usepackage{enumitem} 

\newcommand{\hMsun}{{\ifmmode{h^{-1}{\rm
        {M_{\odot}}}}\else{$h^{-1}{\rm{M_{\odot}}}$~}\fi}} 
\newcommand{\hMpc}{{\ifmmode{h^{-1}{\rm Mpc}}\else{$h^{-1}$Mpc }\fi}}
\def\be{\begin{equation}}
\def\ee{\end{equation}}
\def\ba{\begin{eqnarray}}
\def\ea{\end{eqnarray}}

\received{November 6, 2017}
\revised{November 27, 2017}
\accepted{\today}
\submitjournal{ApJ}

\shorttitle{Non-parametric DE using tomographic AP}

\begin{document}

\title{Non-parametric dark energy reconstruction using the tomographic Alcock-Paczynski test}

\author{Zhenyu Zhang}
\affiliation{School of Physics and Astronomy, Sun Yat-Sen University, Guangzhou 510297, P.R.China}

\author{Gan Gu}
\affiliation{School of Physics and Astronomy, Sun Yat-Sen University, Guangzhou 510297, P.R.China}

\author{Xiaoma Wang}
\affiliation{School of Physics and Astronomy, Sun Yat-Sen University, Guangzhou 510297, P.R.China}

\author{Yun-He Li}
\affiliation{Department of Physics, College of Sciences, Northeastern University, Shenyang 110004, China}

\author{Cristiano G. Sabiu}
\affiliation{Department of Astronomy, Yonsei University, Seoul, Korea}

\author{Hyunbae Park}
\affiliation{Kavli Institute for the Physics and Mathematics of the Universe, University
of Tokyo, Chiba 277-8582, Japan}

\author{Haitao Miao}
\author{Xiaolin Luo}
\author{Feng Fang} 
\author{Xiao-Dong~Li}
\affiliation{School of Physics and Astronomy, Sun Yat-Sen University, Guangzhou 510297, P.R.China}

\email{Corresponding Authors:  lixiaod25@mail.sysu.edu.cn}



\begin{abstract}
The tomographic Alcock-Paczynski (AP) method can result in tight cosmological constraints 
by using small and intermediate clustering scales of the large scale structure (LSS) of the galaxy distribution. 
By focusing on the redshift dependence, the AP distortion can be distinguished from the distortions produced by the redshift space distortions (RSD).
In this work, we combine the tomographic AP method with other recent observational datasets of SNIa+BAO+CMB+$H_0$
to reconstruct the dark energy equation-of-state $w$ in a non-parametric form.
The result favors a dynamical DE at $z\lesssim1$,
and shows a mild deviation ($\lesssim2\sigma$) from $w=-1$ at $z=0.5-0.7$.
We find the addition of the AP method improves the low redshift ($z\lesssim0.7$) constraint by $\sim50\%$.
\end{abstract}




\section{Introduction}

The late-time accelerated expansion of the Universe \citep{Riess1998,Perl1999} implies
either the existence of ``dark energy'' or the breakdown of general relativity on cosmological scales.
The theoretical origin and observational measurements of cosmic acceleration, although have attracted tremendous attention,
are still far from being well explained or accurately measured \citep{SW1989,Li2011,2012IJMPD..2130002Y,DHW2013}.

The Alcock-Paczynski (AP) test \citep{AP1979} enables us to probe the angular diameter distance $D_A$ and the Hubble factor $H$,
which can be used to place constraints on cosmological parameters.
Under a certain cosmological model, the radial and tangential sizes of some distant objects or structures take the forms of 
$\Delta r_{\parallel} = \frac{c}{H(z)}\Delta z$ and $\Delta r_{\bot}=(1+z)D_A(z)\Delta \theta$,
where $\Delta z$, $\Delta \theta$ are their redshift span and angular size, respectively.
Thus, if incorrect cosmological models are assumed for transforming redshifts into comoving distances,
the wrongly estimated $\Delta r_{\parallel}$ and $\Delta r_{\bot}$ induce a geometric distortion, 
known as the AP distortion. 
Statistical methods which probe and quantify the AP distortion
has been developed and applied to a number of galaxy redshift surveys to constrain the cosmological parameters
\citep{Ryden1995,Ballinger1996,Matsubara1996,Outram2004,Blake2011,LavausWandelt2012,Alam2016,Qingqing2016,Doogesh2018}.

Recently, a novel tomographic AP method based on the redshift evolution of the AP distortion
has achieved significantly strong constraints on the cosmic expansion history parameters \citep{topology,Li2014,Li2015,Li2016}.
The method focuses on the redshift dependence to differentiate the AP effect from the distortions produced by the redshift space distortions (RSD),
and has proved to be successful in dealing with galaxy clustering on relatively small scales.
\cite{Li2016} firstly applied the method to the 
SDSS (Sloan Digital Sky Survey) BOSS (Baryon Oscillation Spectroscopic Survey) DR12 galaxies,
and achieves $\sim35\%$ improvements in the constraints on $\Omega_m$ and $w$ 
when combining the method with external datasets of 
the Cosmic Microwave Background (CMB), type Ia supernovae (SNIa), baryon acoustic oscillations (BAO), and the  $H_0$.

In this work we aim to study how the tomographic AP method can be optimised to aid in measuring and characterising dark energy. 
The non-parametric strategy is particularly suitable for constraining functions whose forms are not clearly 
known from the theoretical aspect \citep{Marco2019}.
We apply the method to reconstruct the dark energy equation-of-state $w(z)$,
using the non-parametric approach developed in \cite{Crittenden2009,Crittenden2012,zhao2012},
which has the advantage of not assuming any {\it ad hoc} form of $w$.
In a recent work \cite{ZhaoGB:2017} use this method to 
reconstruct $w$ from 16 observational datasets,
and claim a $3.5\sigma$ significance level in preference of a dynamical dark energy.
It would be interesting to see what the results would be if the tomographic AP 
method is used to reconstruct $w$, 
and whether the reconstructed $w$ is consistent with the results of \cite{ZhaoGB:2017}.

The brief outline of this paper is as follows. 
In \S\ref{sec:method} we outline the tomographic AP method and how we practically implement the non-parametric modelling of $w(z)$.
In \S\ref{sec:results} we present the results of our analysis in combination with other datasets. We conclude in \S\ref{sec:conclusion}. 

\section{Methodology}
\label{sec:method}

In pursuit of reconstructing DE in a model-independent manner,
we adopt the non-parametric method of $w$ \citep{Crittenden2009,zhao2012}
without choosing any particular parameterization.
To start, $w$ is parameterized in terms of its values at discrete steps in the scale factor $a$. 
Fitting a large number of uncorrelated bins would lead to extremely large uncertainties and, 
in fact, would prevent the Monte Carlo Markov Chains (MCMC) from converging due to the large number of degenerate directions in the parameter space. 
On the other hand, fitting only a few bins usually
lead to an unphysical discrete distribution of $w$
and significantly bias the result. 
The solution is to introduce a prior covariance among a large number of bins 
based on a phenomenological two-point function,
\begin{equation}
\xi_w(|a-a^\prime|) \equiv \left<  [w(a)-w^{\rm fid}(a)][w(a^\prime)-
w^{\rm fid}(a^\prime)] \right>,
\end{equation}
which is chosen as the form of \citep{Crittenden2009}, 
\begin{equation}\label{eq:CPZ}
\xi_{\rm CPZ}(\delta a) =  \xi_w (a=0) /[1 + (\delta a/a_c)^2],
\end{equation}
where $\delta a\equiv|a-a^\prime|$. 
Clearly, $a_c$ describes the typical smoothing scale,
and $\xi_w(0)$ is the normalization factor determined by 
the expected variance of the mean of the $w$'s, $\sigma^2_{\bar{w}}$. 
The `floating' fiducial is defined as the local average,
\begin{equation}
w^{\rm fid}_i = \sum_{|a_j - a_i| \leq a_c} w^{\rm true}_j / N_j,
\end{equation}
where $N_j$ is the number of neighbouring bins lying around the $i$-th bin 
within the smoothing scale.

In practice, one should set the priors to conduct the analysis.
A very weak prior (i.e., small $a_c$ or large $\sigma^2_{\bar w}$)
can match the true model on average (i.e., unbiased),
but will result in a noisy reconstruction.
A stronger prior reduces the variance but pulls the reconstructed results towards the peak of the prior.
In this paper, we use the ``weak prior''
$a_c=0.06$, $\sigma_{\bar w}=0.04$, the prior which was also adopted in \citet{zhao2012}.
The tests performed in \cite{Crittenden2009} shown that the results are largely independent of the choice of the correlation function. 
Also, \cite{Crittenden2011} has showed that 
a stronger prior $\sigma_{\bar w}=0.02$ is already enough 
for reconstructing a range of models without introducing a sizeable bias.

We parametrize $w$ in terms of its values at $N$ points in $a$,
i.e., 
\begin{equation}
w_i=w(a_i),\ i=1,2,...,N.
\end{equation}
In this analysis we choose $N=30$,
where the first 29 bins are uniform in $a\in[0.286,1]$, corresponding to $z\in[0,2.5]$,
and the last bin covers the wide range of $z\in[2.5,1100]$.
Given the binning scheme, together with the covariance matrix $\bf C$ given by Equation \ref{eq:CPZ},
it is straightforward to write down prior following the Gaussian form PDF
\begin{equation}
\mathcal{P}_{\rm prior}(\bf w) \propto exp\left( -\frac{1}{2}(\bf{w}-\bf{w^{\rm fid})}
\bf{C}^{-1} (\bf{w}-\bf{w^{\rm fid}} ) \right).
\end{equation}
Effectively, the prior results in a new contribution to the total likelihood of the model given the datasets $D$, 
\begin{equation}
{\cal P}({\bf w}|{\bf D}) \propto {\cal P}({\bf D}|{\bf w}) \times {\cal P}_{\rm prior}({\bf w}),
\end{equation}
thus penalizes those models who are less smooth.

The method is then applied to a joint dataset of recent cosmological observations
including the CMB temperature and polarization anisotropies measured by full-mission Planck \citep{Planck2015}, 
the ``JLA'' SNIa sample \citep{JLA},
a Hubble Space Telescope measurement of $H_0=70.6\pm3.3$ km/s/Mpc \citep{Riess2011,E14H0},
and the BAO distance priors measured from 6dFGS \citep{6dFGS},
SDSS MGS \citep{MGS}, and the SDSS-III BOSS DR11 anisotropic measurements \citep{Anderson2013},
as was also adopted in \citet{Li2016,Li2018}.

These datasets are then combined with the AP likelihood of SDSS-III BOSS DR12 galaxies \citep{Li2016,Li2018},
for which we evaluate the redshift evolution of LSS distortion induced by wrong cosmological parameters
via the anisotropic correlation function,
\begin{equation}\label{eq:deltahatxi}
\delta \hat\xi_{\Delta s}(z_i,z_j,\mu)\ \equiv\ \hat\xi_{\Delta s}(z_i,\mu) - \hat\xi_{\Delta s}(z_j,\mu).
\end{equation}
$\xi_{\Delta s}(z_i,\mu)$ is the integrated correlation function which captures the information of 
LSS distortion within the clustering scales one were interested in,
\begin{equation}
\xi_{\Delta s} (\mu) \equiv \int_{s_{\rm min}=6\ h^{-1}\ \rm{Mpc}}^{s_{\rm max}=40\ h^{-1}\ \rm{Mpc}} \xi (s,\mu)\ ds.
\end{equation}
It was then normalized to remove the uncertainty from clustering magnitude and the galaxy bias,
\begin{equation}
\hat\xi_{\Delta s}(\mu) \equiv \frac{\xi_{\Delta s}(\mu)}{\int_{0}^{\mu_{\rm max}}\xi_{\Delta s}(\mu)\ d\mu}.
\end{equation}
As described in Equation \ref{eq:deltahatxi},
the difference between $\hat\xi_{\Delta s}(\mu)$ measured at two different redshifts $z_i,\ z_j$ 
characterizes the amount of the redshift evolution of LSS distortion.
SDSS DR12 has 361\,759 LOWZ galaxies at $0.15<z <0.43$, 
and 771\,567 CMASS galaxies at $0.43< z < 0.693$.
We split these galaxies into six, non-overlapping redshift bins of
$0.150<z_1<0.274<z_2<0.351<z_3<0.430<z_4<0.511<z_5<0.572<z_6<0.693$
\footnote{The boundaries are determined so that, for LOWZ and CMASS samples, the number of galaxies are same in each bin, respectively.} \citep{Li2016}.

\cite{Li2014,Li2015} demonstrated that $\delta \hat\xi_{\Delta s}(z_i,z_j,\mu)$ is dominated by the AP distortion 
while being rather insensitive to the RSD distortion,
enabling us to avoid the large contamination from the latter and probe the AP distortion information on relative small clustering scales. 

The only difference in our treatment from \cite{Li2016} is that
here we slightly improve the method and adopt a ``full-covariance matrix'' likelihood
\begin{equation}\label{eq:chisq2}
{\cal P}_{\rm AP}({\bf w}|{\bf D}) \propto  \exp\left( -\frac{1}{2}\ {\bf \theta}_{\rm AP}\ \bf{C}_{\rm AP}^{-1}\ {\bf \theta}_{\rm AP}\right ),
\end{equation}
where the vector
\begin{equation}
{\bf \theta}_{\rm AP} = \left[ \hat\xi_{\Delta s}(z_2,z_1,\mu_j),\hat\xi_{\Delta s}(z_3,z_2,\mu_j),.., \hat\xi_{\Delta s}(z_6,z_5,\mu_j)\right]
\end{equation}
summarizes the redshift evolution among the six redshift bins into its
$5\times n_{\mu}$ components ($n_\mu$ is the number of binning in $\xi_{\Delta s}$).
The covariance matrix ${\bf C}_{\rm AP}$ is estimated using the 2,000 MultiDark-Patchy mocks \citep{MDPATCHY}.
Compared with \cite{Li2016}, where the 1st redshift bin is taken as the reference, 
this current approach includes the statistical uncertainties in the system 
and avoids the particular dependence on which specific redshift bin is chosen as the reference.

This improved methodology was presented in \cite{Li2019},
where the authors detailedly explained how the multi-redshifts correlation is included,
and how it affects the constraints on the various cosmological parameters.



\section{Results}\label{sec:results}

\begin{figure*}
   \centering{
   \includegraphics[width=12cm]{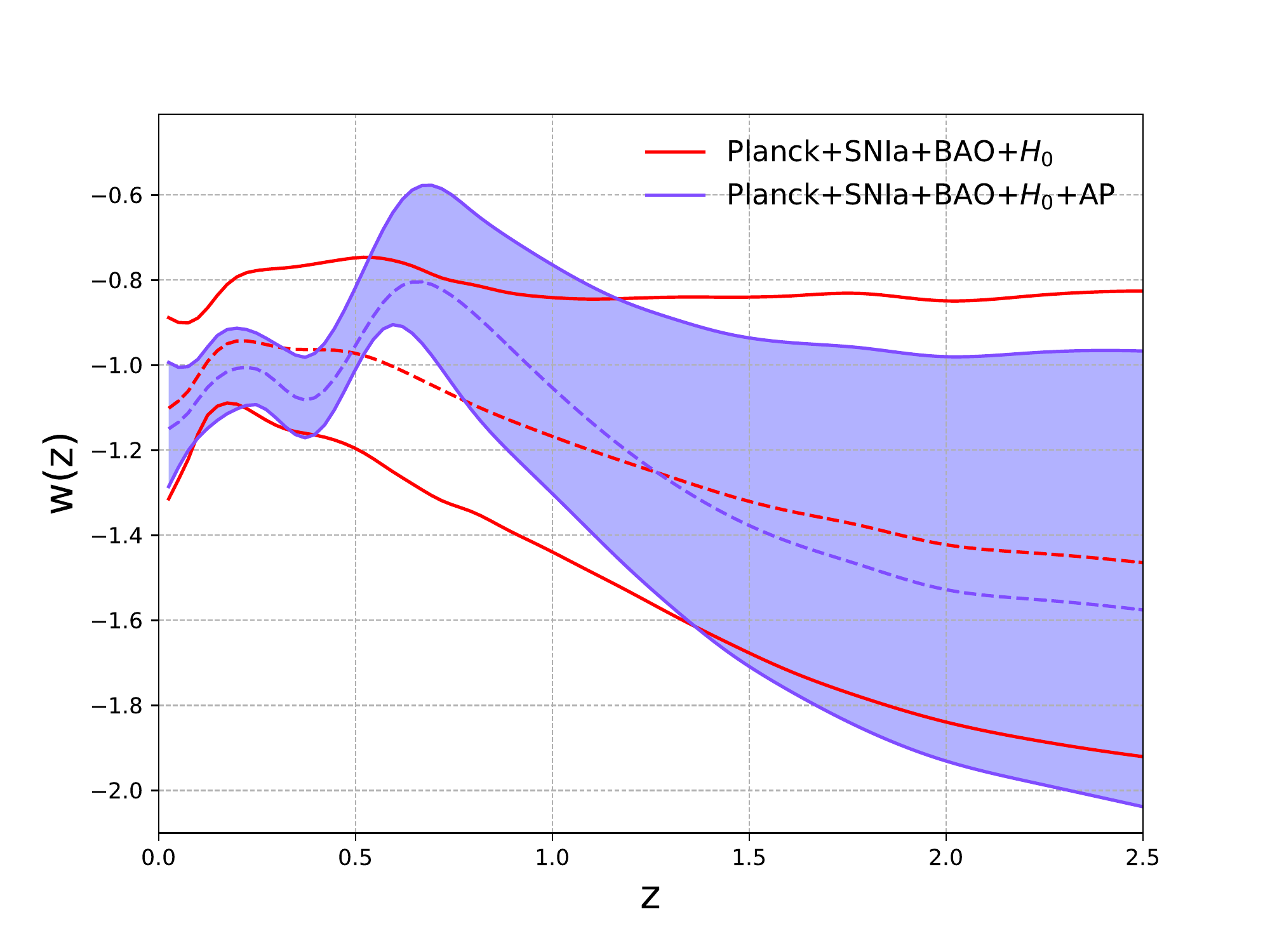}
   }
   \caption{\label{fig_wz}
   Derived redshift evolution of $w(z)$. The mean values and 68.3\%  CL regions are plotted.
   Adding the AP method tightens the constraints.
   Dynamical behavior of dark energy is mildly favored at $z\lesssim0.7$.
   }
\end{figure*}

The derived constraints on $w$ as a function of redshift are plotted in Figure \ref{fig_wz}.
The red solid lines represent the 68.3\% CL constraints based on Planck+SNIa+BAO+$H_0$,
while the AP-added results are plotted in blue filled
\footnote{ The AP and BAO methods probe galaxy clustering on very different scales,
so it is safe to assume they have no correlation and can be simply combined.
In \cite{zhangxue2018}, the authors have computed the 
the correlation coefficient of the anisotropic information in the clustering scales of AP and BAO methods in N-body simulations,
and find it as small as $-0.054\pm 0.034$.}.

The reconstructed $w(z)$ from  Planck+SNIa+BAO+$H_0$
\footnote{ Note that the SDSS DR11 anisotropic BAO measurements also contains the AP information on scales of $\sim 100 h^{-1}$ Mpc,
so it is a little inappropriate to use the abbreviations ``BAO'' and ``AP'' in the legend.
Anyway, we still use them, and our ``AP'' only stands for the 6-40 $h^{-1}$ Mpc tomographic AP measurements of SDSS DR12 galaxies.}
is fully consistent with 
the cosmological constant;
the $w=-1$ line lies within the 68.3\% CL region.
In the plotted redshift range ($0<z<2.5$), 
the upper bound of $w$ is constrained to $\lesssim-0.8$, 
while the lower bound varies from -1.3 at $z=0$ to -2.0 at $z\gtrsim2$, 
dependent on the redshift.
The best constrained epoch lies around $z=0.2$.
These features are consistent with the previous results presented
in the literature using a similar dataset \citep{Zhao:2017cud}.

The constraints are much improved after adding AP to the combined dataset.
At $z\lesssim 0.7$, 
i.e. the redshift range of the SDSS galaxies analyzed by the AP method,
the uncertainty of $w(z)$ is reduced by $\sim$50\%,
reaching as small as 0.2.
It then increases to 0.4-1.0 at higher redshift ($0.7<z<2.5$).
This highlights the power of the AP method in constraining 
the properties of dark energy, 
which were shown in \cite{Li2016,Li2018}.

Although here the AP method only probes the expansion history information at $z\in(0.1, 0.7)$,
it can still affect the high redshift constraints.
At $0.7\lesssim z\lesssim1.0$, the constraints are tightened by the correlated prior of $w(z)$.
At higher redshift, the error bars are less affected,
but the values of $w$ are shifted to more negative regions.
This is due to the combination of AP and CMB data.
Effectively, the CMB data constrain the $w(z)$ in a manner of the ``shift parameter''
$R\equiv\sqrt{\Omega_m}H_0(1+z_*)D_A(z_*)$ \citep{Bond1997},
which constraints the integration of $1/H(z)$ in the range of $z\approx0-1100$.
So if the constraints on $z\lesssim1.0$ are changed, 
the $z\gtrsim1.0$ parts are also altered, correspondingly.

The most interesting phenomenon from our studies is that 
the result indicates a mild discrepancy with a constant $w=-1$.
At $0.5\lesssim z\lesssim0.7$, $w>-1$ is slightly favored ($\lesssim2\sigma$).
The statistical significance of this result is not large enough to claim a detection of deviation from a cosmological constant, 
however this may be readdressed in the near future as the constraining power will become much improved 
when combining tomographic AP with the upcoming experiments of DESI \citep{DESI} or EUCLID \citep{EUCLID}.

The results also slightly favor a dynamical behavior of DE.
At $z=0-0.5$, we find phantom-like dark energy $-1.2\lesssim w \lesssim-1.0$,
while at higher redshift $z=0.5-0.7$ it becomes quintessence-like, $-1.0 \lesssim w \lesssim -0.6$. 
Theoretically, this is known as the quintom dark energy \citep{quintom1}.

The advantage of the tomographic AP method is that,
it makes use of the clustering information in a series of redshift bins
(rather than compresses the whole sample into a single effective redshift).
Thus, it is able to capture the dynamical behavior of dark energy
within narrow ranges of $\Delta z$.

Our results are consistent with the $w(z)$ obtained in \cite{Li2018},
where the authors used the Planck+SNIa+BAO+$H_0$+AP dataset 
to constrain the CPL parametrization $w=w_0+w_a \frac{z}{1+z}$.
They found 100\% improvement in the DE figure-of-merit and 
a slight preference of dynamical dark energy.
Benefitting from a more general form of a non-parameteric  $w(z)$,
we are able to obtain more detailed features in the reconstruction.

\begin{figure*}
   \centerline{
   \includegraphics[width=12cm]{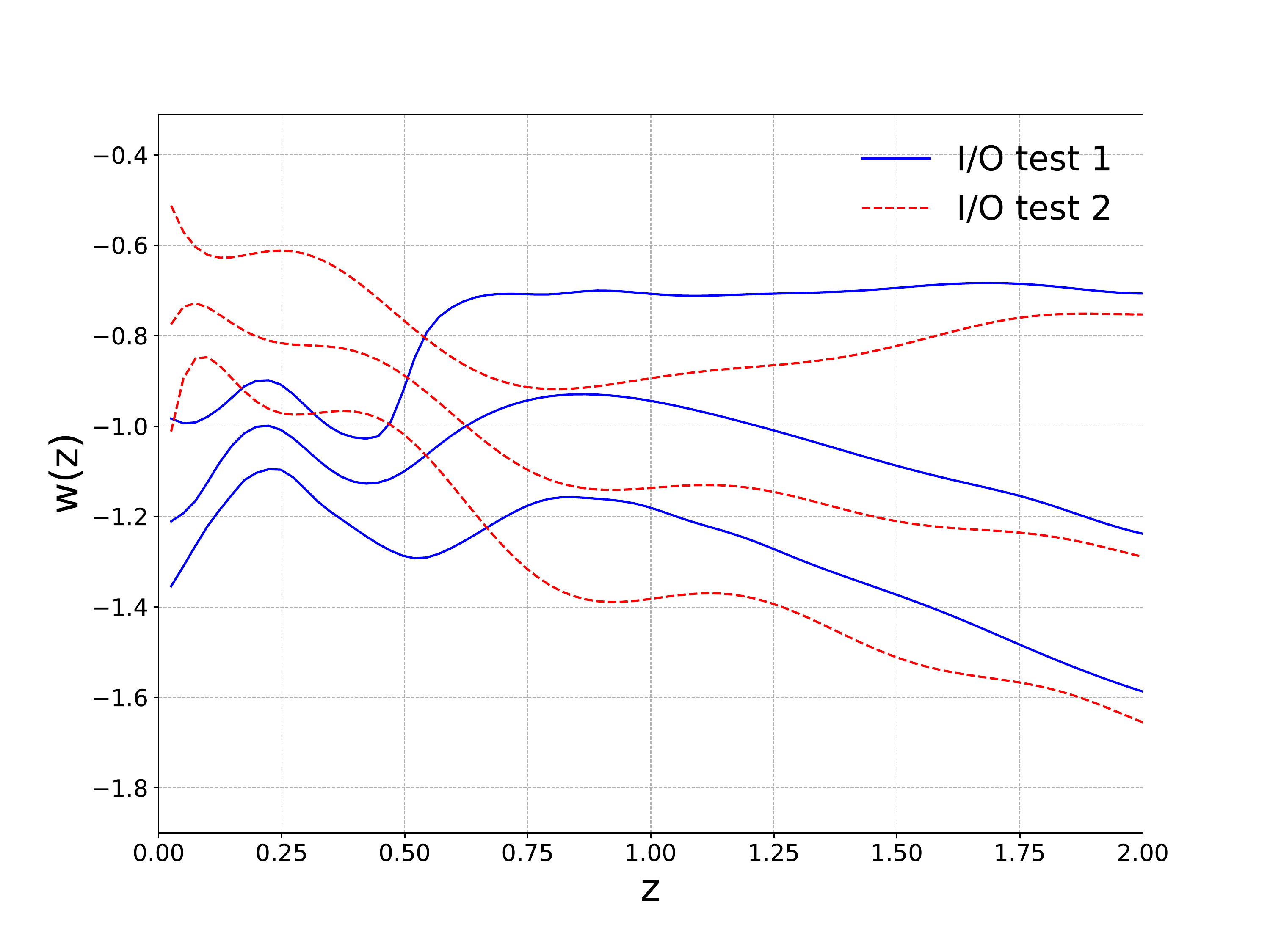}
   }
   \caption{\label{fig_IOpdf}
     Input-output (I/O) test of the tomographic AP method, based on two MultiDark-Patchy realizations.
    Using the method the input cosmology $w=-1$ can be well recovered.
   }
\end{figure*}

 To further validate the results, we did an input-output test on two MultiDark-Patchy realizations
\footnote{ As a simple check we just did it on two realizations. 
Due to the many degrees of freedom the MCMC chains converge very slowly, 
making it rather difficult to perform this kind of test on a large number of mocks.}.
We treat the mocks as the  ``real data'' and apply the AP method to them.
The constraints on $w(z)$ are plotted in Figure \ref{fig_IOpdf}.
The ``true'' cosmology of $w=-1$ are nicely recovered (deviation $\lesssim 1\sigma$).
The size of error bars are $\sim 1.5$ times of the Planck+SNIa+BAO+$H_0$+AP contraints of Figure 
\ref{fig_wz}.
This justifies the ability of the tomographic AP method in 
constraining the non-parametric dark energy equation of state.

Finally, we note that the results with and without AP are in good consistency with each other.
This implies that the information obtained from the AP effect 
agrees well with the other probes.
Since the clustering information probed by AP
is independent from those probed by BAO (see the discussion in \cite{zhangxue2018}),
to some extent,
in this analysis these two  different LSS probes compliment and validate each other.
This is also consistent with the results of  \cite{Li2016},
where we found the contour region constrained by AP consistently overlaps with
those of SNIa, BAO and CMB.

\begin{figure*}
   \centering{
   \includegraphics[width=16cm]{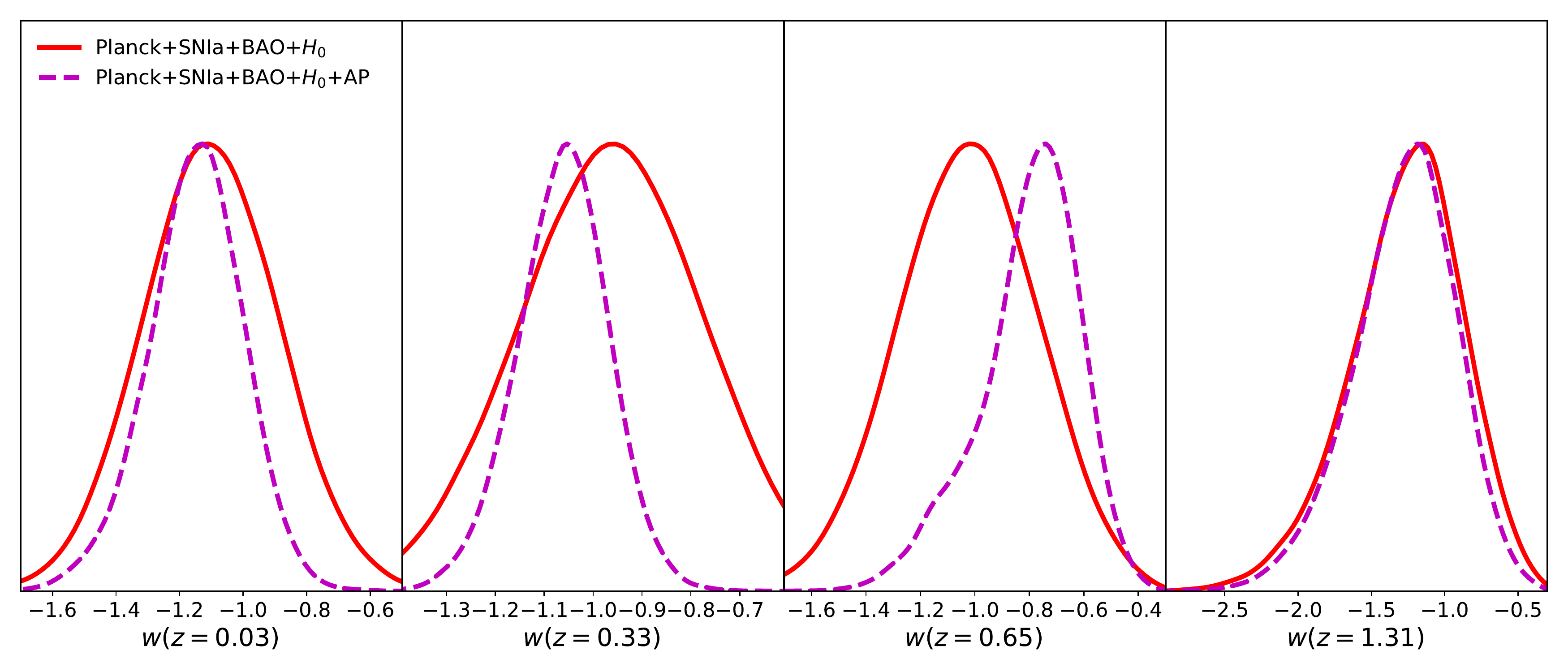}
   }
   \caption{\label{fig_wpdf}
    Likelihoods of $w(z=0.03, 0.33, 0.65, 1.31)$
    from Planck+SNIa+BAO+$H_0$ (red solid) 
    and Planck+SNIa+BAO+$H_0$+ AP(purple dotted). 
    The addition of AP improves the constraint at $z\lesssim 0.7$ by $\sim50$\%.
   }
\end{figure*}

\section{Concluding Remarks}\label{sec:conclusion}

In this work, we consider a very general, non-parametric form for the evolution of the dark energy equation-of-state, $w(z)$.
We obtain cosmological constraints by combining our tomographic AP method with other recent observational datasets of SNIa+BAO+CMB+$H_0$.
As a result, we find that the inclusion of AP improves the low redshift ($z<0.7$) constraint by $\sim50\%$.
Moreover, our result favors a dynamical DE at $z\lesssim1$,
and shows a mild deviation ($\lesssim2\sigma$) from $w=-1$ at $z=0.5-0.7$.

We did not discuss the systematics of the AP method in details.
This topic has been extensively studied in \cite{Li2016,Li2018},
where the authors found that for the current observations the systematical error is still much less than the statistical uncertainty.

We note that our constraint on $w(z)$ at $z\lesssim0.7$ is the tightest within the current literature.
The accuracy  we achieved is as good as that of \citet{Zhao:2017cud} in their  ``ALL16'' combination,
where they used the Planck+SNIa+BAO+$H_0$ datasets\footnote{\cite{Zhao:2017cud} used the SDSS galaxy BAO measurements at nine effective redshifts, 
which are measurements at more redshift points than our adopted BAO dataset, and is expected to be more powerful in such a $w(z)$ reconstruction analysis.},
combined with the WiggleZ galaxy power spectra \citep{Parkinson2012}, 
the CFHTLenS weak lensing shear angular power spectra \cite{Heymans2013},
the $H(z)$ measurement using relative
age of old and passively evolving galaxies
based on a cosmic chronometer approach \citep[OHD; ][]{Moresco2016},
and the Ly$\alpha$ BAO measurements \citep{Deblubac2015}.
In comparison, we use a much smaller number of datasets to achieve a similar low-redshift $w(z)$ constraint.
This highlights the great power of our tomographic AP method using anisotropic clustering on small scales.

At higher redshift ($z\gtrsim0.7$) our constraint is weaker than \citet{Zhao:2017cud}. 
It would be interesting to include more datasets \citep[e.g. the ones used in their paper, the SDSS IV high redshift results,][]{2019MNRAS.482.3497Z}
and then re-perform this analysis.

The dynamical behavior of dark energy at $z\approx0.5-0.7$ has also been found in many other works \citep{Zhao:2017cud,Wang:2018fng}.
Due to the limitation of current observations, 
it is not possible to claim a detection 
of dynamical dark energy at $>5\sigma$ CL.
We expect this can be achieved (or falsified) in the near future aided by more advanced LSS experiments, such as DESI \citep{DESI}, Euclid \citep{EUCLID}, and LSST \citep{LSST}.

\acknowledgments

We thank Gong-bo Zhao, Yuting Wang and Qing-Guo Huang for helpful discussion.
XDL acknowledges the supported from NSFC grant (No. 11803094). 
YHL acknowledges the support of the National Natural Science Foundation of China (Grant No.~11805031) 
and the Fundamental Research Funds for the Central Universities (Grant No.~N170503009). 
CGS  acknowledges  financial  support  from  the  NRF (Grant No.~2017R1D1A1B03034900). 

Based on observations obtained with Planck (\url{http://www.esa.int/Planck}),
an ESA science mission with instruments and contributions directly funded by
ESA Member States, NASA, and Canada.

Funding for SDSS-III has been provided by the Alfred P. Sloan Foundation, the Participating Institutions, the
National Science Foundation, and the U.S. Department of Energy Office of Science. 
The SDSS-III web site is \url{http://www.sdss3.org}. 
SDSS-III is managed by the Astrophysical Research Consortium for the Participating Institutions
of the SDSS-III Collaboration including the University of Arizona, the Brazilian Participation Group, Brookhaven
National Laboratory, Carnegie Mellon University, University of Florida, the French Participation Group, 
the German Participation Group, Harvard University, the Instituto de Astrofisica de Canarias, the Michigan State/Notre
Dame/JINA Participation Group, Johns Hopkins University, Lawrence Berkeley National Laboratory, Max Planck
Institute for Astrophysics, Max Planck Institute for Extraterrestrial Physics, New Mexico State University, New
York University, Ohio State University, Pennsylvania State
University, University of Portsmouth, Princeton University,
the Spanish Participation Group, University of Tokyo, University of Utah, Vanderbilt University, University of Virginia, 
University of Washington, and Yale University.

\end{document}